\documentclass[fleqn,twoside]{article}
\usepackage{espcrc2}

\usepackage{graphicx}
\usepackage[figuresright]{rotating}

\newcommand{\AmS}{{\protect\the\textfont2
  A\kern-.1667em\lower.5ex\hbox{M}\kern-.125emS}}

\title{Optical Counterpart of The Ultraluminous X-ray Source NGC 1313 X-2}

\author{L. Zampieri\address[INAF]{INAF-Osservatorio Astronomico di
Padova, Vicolo dell'Osservatorio 5, I-35122 Padova, Italy},
	P. Mucciarelli\addressmark,
	R. Falomo\addressmark,
	P. Kaaret\address[CfA]{Harvard-Smithsonian Center for Astrophysics, 60 Garden Street,
Cambridge, MA 02138},
	R. Di Stefano\addressmark,
	R. Turolla\address[Fispd]{Dipartimento di Fisica, Universit\`a di Padova, Via Marzolo 8, I-35131
Padova, Italy},
	M. Chieregato\address[Insub]{Dipartimento di Scienze, Universit\`a dell'Insubria, Via Valleggio 11,
I-22100 Como, Italy}
	and
	A. Treves\addressmark.}
       
\begin{document}

\begin{abstract}
We present new optical and {\it Chandra} observations of the field
containing the ultraluminous X-ray source NGC1313 X-2. On an ESO 3.6 m
image, the {\it Chandra} error box embraces a $R=21.6$ mag
stellar-like object and excludes a previously proposed optical
counterpart. The resulting X-ray/optical flux ratio of NGC 1313 X-2 is
$\sim 500$. The value of $f_X/f_{opt}$, the X-ray variability history
and spectral distribution indicate a luminous X-ray binary in NGC 1313
as a likely explanation for NGC 1313 X-2.
The inferred optical luminosity ($L\approx 10^5 L_\odot$) is
consistent with that of a $\approx 10 M_\odot$ companion.
\vspace{1pc}
\end{abstract}

\maketitle

\section{INTRODUCTION}

Point-like, off-nuclear X-ray sources with luminosities significantly
exceeding the Eddington limit for one solar mass are being
progressively discovered in the field of many nearby galaxies
(e.g. \cite{col02}). These powerful objects, commonly referred to as
ultraluminous X-ray sources (ULXs), do not appear to have an obvious
Galactic counterpart. Despite some of them have been identified with
supernovae or background active galactic nuclei, the nature of most of
these sources remains unclear. Among the various possibilities, the
most favored explanation is that ULXs are powered by accretion and
that they are somewhat special X-ray binaries, either containing an
intermediate mass black hole (BH) with $M_{BH}\geq 100
\, M_\odot$ (see e.g. \cite{colbert99,kaaret01}) or having beamed
emission toward us (see e.g. \cite{kaaret03}). For a recent review on
the properties of ULXs we refer to \cite{fabbiano03}. Optical
observations are of fundamental importance to better assess the nature
of these sources but they are still rather scarce (see
e.g. \cite{cagn03,fosc02b}). A certain number of ULXs have optical
counterparts in the Digitized Sky Survey or Hubble Space Telescope
images (e.g. NGC 5204 X-1 \cite{rob01,goad02}) and some appear to be
embedded in emission nebulae a few hundred parsecs in diameter
\cite{pak02}.

NGC1313 X-2 was one of the first sources of this type to be found. It
was serendipitously discovered in an {\it Einstein\/} IPC pointing
toward the nearby SBc galaxy NGC 1313 \cite{fab87}. Originally
included in the {\it Einstein\/} Extended Medium Sensitivity Survey as
MS 0317.7-6647, it is located $\sim 6'$ south of the nucleus of NGC
1313. Stocke et al. \cite{sto95} performed multi-wavelength
observations of MS 0317.7-6647, identified a possible optical
counterpart and proposed that the source could be either a Galactic
isolated neutron star or a binary containing a massive BH in NGC 1313.
A very recent analysis of a {\it XMM\/} EPIC MOS observation of NGC
1313 \cite{mil02} indicates that two spectral components, soft and
hard, are required to fit the spectrum of NGC 1313 X-2 and that the
normalization of the soft component yields $M_{BH}\geq 830 M_\odot$.

We present new optical\footnote{Based on observations collected at the
European Southern Observatory, Chile, Program number 68.B-0083(A).} 
and {\it Chandra} observations of NGC 1313 X-2, with the aim to shed
further light on its enigmatic nature\footnote{Results reported in
these Proceedings are based on new observations and differ in part
from those originally presented at the Meeting.}.

\section{X-RAY AND OPTICAL DATA}

\subsection{X-ray astrometry}

NCG 1313 X-2 was observed by {\it Chandra\/} on 13 Oct 2002.  The
observation had a duration of 19.9 ks.  The primary goal of the
observation was to study sources near the center of the galaxy, but
the aim-point was adjusted to also place NGC 1313 X-1, NGC 1313 X-2,
and SN 1978K on the S3 chip of the ACIS-S. Data were extracted and
subjected to standard processing and event screening. No strong
background flares were found, so the entire observation was used.
Because the source is 5$'$ off axis, the point spread function was
fitted with an ellipsoidal Gaussian (1.9$''$ along the major axis and
1.1$''$ along the minor axis, rms values).  Also, the pixel with the
highest number of counts is offset by 0.8$''$ from the center of the
fitted ellipse. Taking these uncertainties into account, we
conservatively estimate a positional error of 0.7$''$ (1-$\sigma$).
The final {\it Chandra} position is: $\alpha=$ 03h 18m
22.27s$\pm$0.12s, $\delta=$ -66$^0$ 36$'$ 03.8$'' \pm$0.7$''$.

In order to check the accuracy of the {\it Chandra} aspect solution,
we exploited the presence in the field of view of a quite peculiar
supernova, SN 1978K, that shows powerful radio and X-ray emission. The
{\it Chandra} position of SN 1978K is $\alpha=$ 03h 17m 38.69s,
$\delta=$ -66$^0$ 33$'$ 03.6$''$ (J2000), within $0.46''$ from the
accurate (0.1$''$) radio position of \cite{ryd93}. This is consistent
with the expected {\it Chandra} aspect accuracy.

The position of NGC 1313 X-2 was previously determined from a {\it
ROSAT\/} HRI \cite{sto95,sch00} and a {\it XMM} EPIC-MOS
\cite{mil02} observation. Typical 1-$\sigma$ error boxes for both
instruments are $\sim 3''$ for {\it ROSAT\/} HRI and $\sim 2''$ for
{\it XMM} EPIC-MOS. The {\it ROSAT\/} and {\it XMM} positions and
corresponding error boxes are summarized in Table \ref{tab0}.

\begin{table*}[htb]
\caption{Positions of NGC1313 X-2 and
positions and optical magnitudes of field objects.}
\label{tab0}
\newcommand{\m}{\hphantom{$-$}}
\newcommand{\cc}[1]{\multicolumn{1}{c}{#1}}
\renewcommand{\arraystretch}{1.2} 
\begin{tabular}{@{}llllll}
\hline
Observatory/Instr. & Object$^a$ & RA[J2000] & DEC[J2000] & R magnitude &
Ref. \\
\hline
{\it ROSAT\/}/HRI     & NGC1313 X-2     &  03 18 22.00$\pm$0.50
& -66 36 02.3$\pm$3.0   &       --&     \cite{sch00}\\
{\it XMM\/}/EPIC-MOS  & NGC1313 X-2     &  03 18 22.34$\pm$0.33 &
-66 36 03.7$\pm$2.0     &       --&     \cite{mil02}\\
{\it Chandra\/}/ACIS-S& NGC1313 X-2     &  03 18 22.18$\pm$0.12 &
-66 36 03.3$\pm$0.7     &       --&     this work   \\
ESO/3.6m      &  A    & 03 18 21.97$\pm$0.05  & -66 36 06.5$\pm$0.3 &
19.8$\pm$0.2 &
this work\\
ESO/3.6m      &  B    & 03 18 21.56$\pm$0.05  & -66 36 00.9$\pm$0.3 & 
20.7$\pm$0.2  &
this work\\
ESO/3.6m      &  C    & 03 18 22.34$\pm$0.05  & -66 36 03.7$\pm$0.3 &
21.6$\pm$0.2 &
this work\\
ESO/3.6m      &  D    & 03 18 20.96$\pm$0.05  & -66 36 03.7$\pm$0.3 &
17.8$\pm$0.2 &
this work\\
\hline
\end{tabular}\\[2pt]
$^a$ See Figure~\ref{fig3}.
\end{table*}

\subsection{Optical astrometry and photometry}

Optical images of the field of NGC1313 X-2 in the $R$-band (Bessel
filter) were taken on 16 January 2002 with the 3.6 m telescope of the
European Southern Observatory (ESO) at La Silla (Chile). We used
EFOSC2 with a Loral/Lesser CCD of 2048$\times$2048 pixels yielding a
field of view of $\sim 5' \times 5'$ at a resolution of
0.314$''$/pixel (re-binned by a factor 2). The night was clear with a
seeing of about 1$''$.  Four images were obtained for a total exposure
time of 1320 s. Standard reduction of the data (including bias
subtraction and flat-field correction) was performed within the IRAF
(v 2.12) environment.

Our four ESO images were astrometrically calibrated using an IRAF task
({\sc PLTSOL}) and performing a polynomial interpolation starting from
the positions of GSC2 ESO field stars. The internal accuracy of this
procedure was estimated comparing the actual positions of a number of
GSC2 stars not used for astrometric calibration with the positions
contained in the catalog. The accuracy is 0.3" (1-$\sigma$). The four
calibrated images were then summed together and the resulting image is
shown in Figure \ref{fig3}.

In order to check for the relative systematics between the optical and
X-ray astrometric calibrations, we used the position of SN 1978K. This
supernova is inside the {\it Chandra} field of view but outside our
optical image. Thus, we analyzed also an archival image of SN 1978K
(from the Padova-Asiago Supernova Archive) taken on 13 September 1999
with the same telescope and a similar instrumental set-up (ESO
3.6m+EFOSC/2.9+R\#642, exposure time 180 s). After calibrating the
archival image, the position of SN 1978K is $\alpha=$ 03h 17m 38.605s,
$\delta=$ -66$^0$ 33$'$ 03.13$''$ (J2000). This is within 0.28$''$
from the radio position of \cite{ryd93}, improving significantly upon
the previous optical position by the same authors. The difference
between the centroids of the optical and {\it Chandra} positions of SN
1978K is 0.69$''$ ($\alpha_{opt}-\alpha_X=$ -0.085s,
$\delta_{opt}-\delta_X=$ -0.47$''$). Although this difference is small
and comparable with the statistical errors, we decided to apply this
correction to the {\it Chandra} position of NGC 1313 X-2 to eliminate
any systematic error between the optical and X-ray astrometric
calibrations. The resulting {\it Chandra} position of NGC 1313 X-2 is
reported in Table \ref{tab0}.

The photometry of the objects in our optical image was performed
calibrating the frame with the $R$-band magnitudes of 23 field stars
from the SuperCosmos Sky Survey. The internal accuracy of this
calibration procedure is 0.2 mag at 1-$\sigma$. The magnitudes of the
relevant objects in the field are reported in Table \ref{tab0}. The
measurement for object D is in good agreement with the value in the
SuperCosmos Sky Survey catalog.

\section{DISCUSSION}

\begin{figure*}[htb]
\vspace{9pt}
\includegraphics[width=14cm]{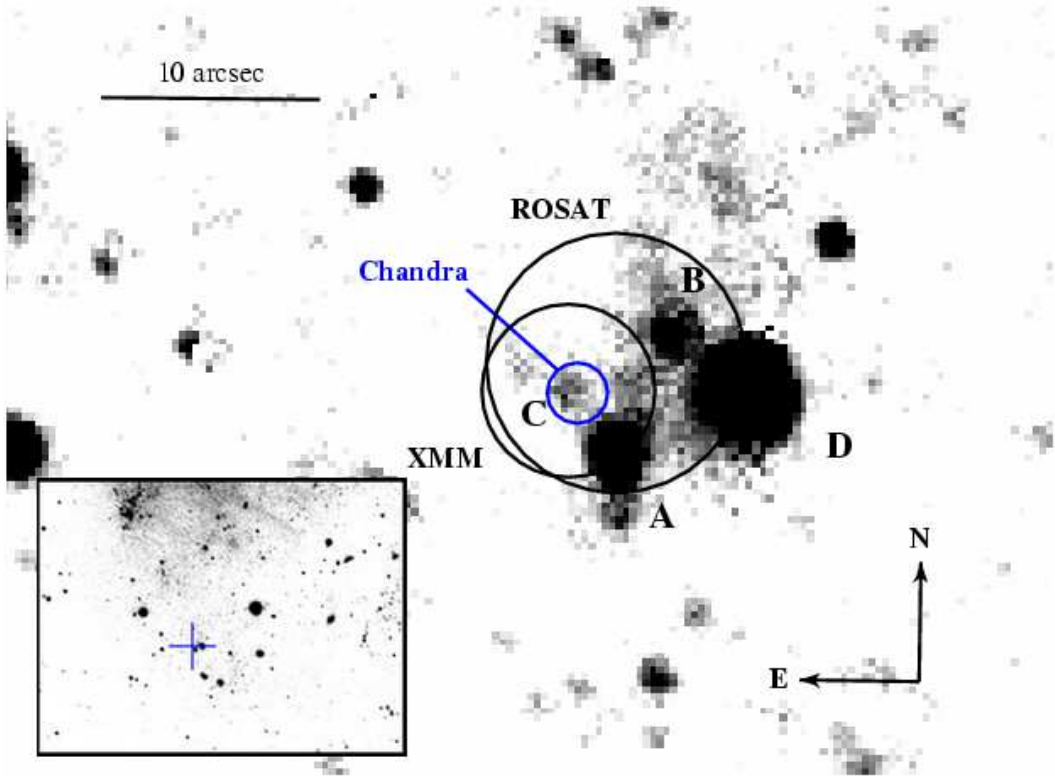}
\caption{ESO 3.6m $R$-band (Bessel filter) image of the field of NGC1313 X-2.
The circles show the {\it ROSAT\/} HRI, {\it XMM\/} EPIC-MOS and {\it
Chandra} ACIS-S positions. The estimated 90\% confidence radii are
6$''$ for HRI, 4$''$ for EPIC-MOS and 1.4$''$ for ACIS-S. Labels A, B,
C and D mark the four field objects inside or close to the X-ray error
boxes. The insert at the bottom-left shows a larger portion of the
image with the position of the X-ray source ({\it cross}).}
\label{fig3}
\end{figure*}

The shifted {\it Chandra} position of NGC1313 X-2, corrected to match
the optical and X-ray positions of SN 1978K, is shown in Figure
\ref{fig3}, together with the {\it ROSAT\/} HRI \cite{sch00} and {\it
XMM} EPIC-MOS \cite{mil02} error boxes, overlaid on our ESO image. All
measurements are consistent within 1-$\sigma$. The distance of the
centroids of objects A, B and D with respect to the (shifted) {\it
Chandra} position is 3.6$''$, 4.1$''$ and 7.3$''$, respectively. Even
taking into account the statistical error on the optical positions
(0.3$''$), these three objects can be ruled out at a significance
level of at least 3-$\sigma$. On the other hand, object C is inside
the Chandra error box and its position coincides within 1-$\sigma$
with that of NGC 1313 X-2, making it a likely counterpart.

From the maximum absorbed X-ray flux of NGC 1313 X-2 ($f_X=2\times
10^{-12}$ erg cm$^{-2}$ s$^{-1}$) and optical magnitude of object C
($R=21.6$), we estimate $f_X/f_{opt}\sim 500$. Only Isolated Neutron
Stars (INSs), heavily obscured AGNs and luminous X-ray binaries can
reach such large values of the X-ray/optical flux ratio. INSs are
extreme in this respect, with B$\approx 25$ optical counterparts and
typical X-ray-to-optical flux ratios $\geq 10^5$ (see
e.g. \cite{kapl03,hulleman01} for the counterparts of Anomalous X-ray
Pulsars). The presence of object C in the {\it Chandra} error box
makes this possibility unlikely. Furthermore, known INSs exhibit
different spectral properties with no significant variability. On the
other hand, a heavily obscured AGN is expected to have a rather hard
X-ray spectrum and to emit significantly in the near-infrared (see
e.g. \cite{brusa02}). Since NGC 1313 X-2 is relatively bright in
X-rays, infrared emission would be expected at a higher level
($K\approx 12$), were it an obscured AGN. The lack of any IR
counterpart on a $K$ image of the 2MASS All Sky Image Service down to
a limiting magnitude $K\simeq 14$ (10-$\sigma$) and the softer X-ray
spectrum of NGC 1313 X-2 rule out this possibility. Our accurate {\it
Chandra} position and optical identification essentially leave only a
very luminous X-ray binary in NGC 1313 as a viable option for NGC 1313
X-2. This is in line with the binary nature of ULXs and is consistent
with the observed properties of this source, such as the X-ray
variability and the observed X-ray spectrum, including the presence of
a soft component probably produced by an accretion disk \cite{mil02}.

If indeed NGC 1313 X-2 is a black hole binary, the X-ray spectral
parameters, in particular the temperature and/or the normalization of
the soft component can be used to estimate the BH mass, as already
done by Miller et al. \cite{mil02}. Fitting the soft component with a
multicolor disk (MCD) blackbody model, they find that NGC1313 X-2
contains an intermediate mass BH with a mass in excess of $100
M_\odot$.
The large inferred BH mass does not require beamed emission. Then, the
estimated accretion rate (assuming 10\% efficiency) is ${\dot M}
\approx 10^{-7} \, M_\odot \, {\rm yr}^{-1}$, forcing the mass reservoir
to be a companion star.

From the apparent magnitude of object C ($R=21.6$) and its visual
absorption ($A_R \simeq 1.6$, computed from the X-ray best fitting
column density $N_H\sim 3\times 10^{21}$ cm$^{-2}$ \cite{bol78}), we
estimate a luminosity $\approx 10^5 L_\odot$, depending on the adopted
bolometric correction. If this originates from the companion star, the
inferred luminosity is consistent with a $\approx 10 M_\odot$
supergiant, making NGC 1313 X-2 a high-mass X-ray binary. Assuming
that 20--30\% of the X-ray flux produced in the innermost part of the
accretion disk intercepts the outer regions, for realistic values of
the albedo ($\geq 0.9$, e.g. \cite{djvna96}), few percents of the
X-ray luminosity ($\approx 10^{38}$ erg s$^{-1}$) can be absorbed and
re-emitted in the optical band. Characteristic emission lines of
X-ray-ionized H, He or N, typically seen in luminous Galactic X-ray
binaries should then be detectable in the optical spectrum. Also X-ray
heating of the companion star itself may contribute to the optical
emission. We can not rule out also that the optical emission detected
in the {\it Chandra} error box originates from a stellar cluster (see
e.g. the case of a ULX in NGC 4565 \cite{wu02}), in which case NGC
1313 X-2 might be a low-mass X-ray binary in the cluster.

The mass accretion rate required to power the observed luminosity may
in principle be provided by Roche-lobe overflow from an evolved
companion or a wind from a supergiant. In the first case, evolutionary
swelling of the companion keeps pace with the increase in Roche lobe
size and the system remains self-sustained: accretion is likely to
proceed through a disk. In the second case, assuming 10\% accretion
efficiency and that the BH can capture $\sim 1\%$ of the mass outflow,
the wind must be very powerful (${\dot M} \sim 10^{-5} \, M_\odot \,
{\rm yr}^{-1}$). A lower efficiency would require too high a gas
supply, making a disk needed even for a wind-fed system. The disk
would probably be much smaller than in a Roche-lobe overflow system
and the optical emission dominated by the supergiant. On the other
hand, in a Roche-lobe overflow system, an extended, possibly
re-irradiated accretion disk should contribute heavily in the UV and
$B$ bands, producing strong emission lines.

A crucial question is how a binary system containing an intermediate
mass BH may have formed. The BH progenitor must have been rather
massive. This is consistent with the fact that NGC 1313 is likely to
have a low metallicity ($Z\sim 0.1-0.2$ \cite{ryd92}). Such a massive
BH may have formed through direct collapse without producing a
supernova. In this way, if the system was born as a binary, it may
have survived after the collapse of the primary. Although less likely,
it is also possible that the companion might have been captured from a
nearby stellar association. In this case, it is not possible to
exclude that the BH may have formed from an early episode of star
formation (population III).

It is worth noting that, although the large BH mass does not require
that the emission is beamed, we cannot rule out that a moderate jet
activity, producing radio emission, is present in NGC 1313 X-2 (see
e.g. the case of an ULX in NGC 5408 \cite{kaaret03}). However,
presently available radio images of the field of NGC 1313 X-2 (Sydney
University Molonglo Sky Survey at 843 MHz and Australia Telescope
Array at $\sim 5$ GHz \cite{sto95}) are not sufficiently deep to
allow detection.

PK acknowledges partial support from NASA grant NAG5-7405 and Chandra
grant GO2-3102X.  This work has been partially supported also by the
Italian Ministry for Education, University and Research (MIUR) under
grant COFIN-2000-MM02C71842 and COFIN-2002-027145.

\end{document}